\documentclass[aps,prb,twocolumn,showpacs,superscriptaddress,floatfix]{revtex4-2}
\usepackage{natbib}
\usepackage{bm}
\usepackage[normalem]{ulem}
\newcommand{\ex}[1]{\langle#1\rangle}
\newcommand{\ti}[0]{\Tilde{i}}

\PassOptionsToPackage{table,dvipsnames}{xcolor}
\usepackage{graphicx}
\usepackage{color}
\usepackage{amsmath}
\usepackage{amssymb}
\usepackage[utf8]{inputenc}
\usepackage{comment}
\usepackage{dcolumn}
\usepackage{bm}
\usepackage{hyperref}
\usepackage{soul}
\usepackage{footmisc}
\usepackage{xfrac}

\usepackage{ulem}
\usepackage{amsfonts}
\usepackage{verbatim}
\usepackage{stackrel}
\usepackage[capitalize]{cleveref}
\usepackage{tikz}
\usetikzlibrary{shapes.geometric, arrows}
\usetikzlibrary{decorations.markings}
\usetikzlibrary{decorations.pathmorphing}
\usetikzlibrary{decorations.pathreplacing}

\graphicspath{{/Images}}
\usepackage{float}
\newcolumntype{P}[1]{>{\centering\arraybackslash}p{#1}}
\definecolor{tdgreen}{rgb}{0,0.4,0}
\definecolor{tlgreen}{rgb}{0,0.7,0}
\definecolor{tpurple}{RGB}{103, 0, 31}
\definecolor{tbrown}{RGB}{127, 39, 4}
\definecolor{tviolet}{RGB}{73, 0, 106}
\definecolor{tgreen}{RGB}{1, 70, 54}
\definecolor{clblue}{rgb}{0.050, 0.35, 0.90}

\newcommand{\Rmnum}[1]{\expandafter\@slowromancap\romannumeral  #1@}

\renewcommand\Im{\operatorname{Im}}
\usepackage{bbold}
\usepackage{color}

\makeatletter
\allowdisplaybreaks
\makeatother

\begin{document}

\title{A Green's function method for the two-dimensional\\
frustrated spin$-1/2$ Heisenberg magnetic lattice}

\author{Zhen Zhao}
\email{Zhen.Zhao@teorfys.lu.se}
 \affiliation{Division of Mathematical Physics and ETSF, Lund University, PO Box 118, 221 00 Lund, Sweden}
 \author{Claudio Verdozzi}
 \email{Claudio.Verdozzi@teorfys.lu.se}
 \affiliation{Division of Mathematical Physics and ETSF, Lund University, PO Box 118, 221 00 Lund, Sweden}
 \author{Ferdi Aryasetiawan}
 \email{Ferdi.Aryasetiawan@teorfys.lu.se}
 \affiliation{Division of Mathematical Physics and ETSF, Lund University, PO Box 118, 221 00 Lund, Sweden}



\date{\today}


\begin{abstract}
The magnon Hedin's equations are derived via the Schwinger functional derivative technique, and the resulting self-consistent Green's function method is used to calculate ground state spin patterns and magnetic structure factors for 2-dimensional magnetic 
systems with frustrated spin-1/2 Heisenberg exchange coupling. Compared to  random-phase approximation treatments, the inclusion of a self-energy correction improves the accuracy in the case of scalar product interactions, as shown by comparisons between our method and exact benchmarks in homogeneous and inhomogeneous finite systems. We also find that for cross-product interactions (e.g. antisymmetric exchange), the method
does not perform equally well, and an inclusion of higher corrections is in order. Aside from indications for future work, our results clearly indicate that the Green's function method in the form proposed here
already shows potential advantages in the description of systems with a large number of atoms as well as long-range interactions.
\end{abstract}

\maketitle
\section{Introduction}
Due to the steady improvement of material fabrication procedures and high-resolution spin-resolved experimental techniques,
in the past few decades the list of novel magnetic phenomena and materials with complex magnetic order  
has grown at fast pace. Notable entries are, for example, magnetoresistance materials \citep{schiffer1995low,ramirez1997colossal}, helical magnets  
\citep{muhlbauer2009skyrmion,yu2010real,seki2012observation,yuan2017skyrmions} and spin liquid systems \citep{balents2010spin,schweika2022chiral},
where interest from fundamental research in novel magnetic behavior merges with aims of technological exploitation in novel electronic devices.

An often distinctive trait of these systems is the occurrence of competing magnetic phases, which can 
change into each other upon a slight change of experimental conditions, and sometimes even exhibit even
re-entrant behavior. This is for example what happens in TlCuCl$_3$ \citep{oosawa2003neutron}, where magnetic order can be tuned by applied pressure, or in the cubic chiral magnet MnSi$_{1-x}$Ge$_x$ \citep{fujishiro2019topological}, 
in which  the spin texture is changed between 
skyrmion lattice and hedgehog lattice on Ge/Si substitution.

A key element in determining the aforementioned variety of complex magnetic behaviors is magnetic frustration, which originates from different and competing magnetic couplings (e.g. spinel cubic materials, like CoAlO$_4$ \citep{suzuki2007melting,lee2008theory}, LiYbO$_2$ \citep{bordelon2021frustrated}) or from specific spin lattice geometries,
in which a given coupling cannot favor one among antagonistic magnetic configurations in a loop around a lattice plaquette \citep{ramirez2003geometric} (as e.g. for pyrochlore-lattice compunds \citep{moessner2001magnets}). 

The notion of spin frustration is of quite general occurrence in magnetism \citep{ramirez1994strongly,moessner1998properties,jiang2009supersolid,balents2010spin,yang2022magnetic}: indeed, it is highly relevant also
for classical degrees of freedom (as e.g. in Ising and Potts models) when, in the large-$S$ limit, 
a classical treatment becomes appropriate. 
However, a quantum description is always in order for systems with spins $S<1$, 
where strong quantum fluctuations are present.  In this case, a customary way to describe 
frustration is via the  
quantum Heisenberg model (QHM), in which spins $S$ 
localized at the nodes of a graph interact via (possibly long ranged) exchange 
interactions. 

The QHM is a popular and flexible conceptual template that can include, among others,
three-spin couplings \citep{baxter1973exact}, multipolar interactions \citep{santini2009multipolar}, chiral spin-interactions \citep{sergienko2006role}, quadratic anisotropy terms \cite{gay1986spin,matsumoto2001ground}, etc. 
Importantly, the QHM is not only useful for its high pedagogical value: 
with a suitable choice of the model parameters (extracted for example by first principle calculations \citep{foyevtsova2011determination,jeschke2013first,liechtenstein1987local}), 
it is often possible to obtain an accurate description of phase diagrams of real materials \citep{matsumoto2001ground}. 

The QHM is also a paradigmatic testground for theoretical methods, and an extensive literature exists on the subject (see  e.g. \citep{manousakis1991spin}). 
Yet, in spite of a vast theoretical effort spanning almost a century, exact solutions are analytically known only in special cases, namely for one-dimensional (1D) chains with only nearest-neigbor (NN) interactions \citep{bethe}, where the Bethe ansatz can be used.
For higher dimensions, or more complicated "flavors" of the QHM, approximations become necessary (as for example, in mean-field level approaches like spin-wave and bosonic methods \citep{arovas1988functional}). 

Exact solutions to the QHM can also be obtained numerically, for example
via exact diagonalization \citep{dagotto1989exact,dagotto1989phase}
(applicable for any dimensionality but limited to very small samples),
via quantum Monte Carlo (QMC) methods (based on stochastic algorithms) \citep{suzuki1977monte,lyklema1982quantum,makivic1991two,
manousakis1991spin,liu1989variational}, and the density matrix renomalisation group (DMRG) \citep{white1992density,white1993density} (originally devised
for 1D systems \citep{white1992density} but recently employed also for higher dimensions \citep{jiang2008density,yan2011spin,stoudenmire2012studying,he2017signatures}).
Fairly large samples can be treated within QMC and DMRG, which are
useful to perform benchmark tests.

Both QMC and DMRG play a central role in our understanding  
large but finite-size clusters and periodic bulk systems with finite/short-range interactions; however, the case of long-range coupling, e.g., dipole interactions, is difficult to address with these approaches, since the range of the interaction can be significantly longer than the 
size of cluster units computationally viable.

A way to overcome this issue is provided by the Green's function (GF) technique. Tyablikov \citep{tyablikov1959retarded} and Kondo \citep{kondo1972green} developed different decoupling methods within the random phase approximation (RPA) to solve the hierarchy problem in the equation of motion of the GF. Similar decoupling methods have been used to study 1- and 2-dimensional $S=1/2$ ferromagnets \citep{yablonskiy1991tyablikov,Junger2004ferro}. 

The GF technique can be applied with relatively low computational load to first-principles treatments of systems with effective spin-dependent interactions \citep{aryasetiawan2008generalized}. In addition, 
the Schwinger functional derivative technique for GF, which goes beyond the RPA decoupling, has been used to calculate the spin-wave spectra \citep{aryasetiawan1999green}. Furthermore, applications of the GF technique are not restricted to cluster models \citep{hamedoun2001quantum}, which makes this technique a good candidate in the study of strongly coherent behavior and the magnetic structure factor of real materials, and in particular to address the case of long-range coupling.
\section{This work, and plan of the paper}
Motivated by these considerations, in this paper we present a new approach to solve a finite $S=1/2$ 2-dimensional frustrated $J_1-J_2$ QHM within the GF scheme. Our  formulation is general and in principle exact but, as usual in practice,
an approximate scheme for vertex corrections beyond the RPA needs to be introduced. Since the magnetic structure factor is closely related to response functions, it is natural to adopt the Schwinger functional derivative technique when deriving the equations for the many-body correlation vertex, which then opens the way to include systematically corrections beyond the RPA.

The main outcomes of our work are:
(i) Derivation of the magnon Hedin's equations for the QHM, which are solved self-consistently within a scheme beyond the RPA; (ii) Inclusion of impurities; (iii) Calculations of the spin correlations functions and the magnetic structure factor for a cluster system using the developed GF method; (iv) Comparisons against exact numerical benchmarks, which show that our approach provides fairly good accuracy with relatively low computational cost.
Most importantly, and more in general, our results suggest that the new method offers
a practical, computationally advantageous route to investigate long-range interactions in the QHM.

The plan of the paper is as follows: in Section 2, we introduce the system and the corresponding Hamiltonian, and derive the magnon Hedin's equations via the Schwinger functional derivative technique; in Section 3, we solve in momentum space the equations self-consistently for 2D lattices ; results and their discussion are presented in Section 4; finally, in Section 5 we provide some conclusive remarks and an outlook.

\section{Theory}\label{theorysect}
The Schwinger functional derivative technique is used here to relate the high-order Green's function (GF) to the response of the lower-order one with respect to a probing field. The vertex equation of GF is derived and is solved self-consistently. The procedure presented here provides a formal justification of Tyablikov's decoupling while at the same time producing improved results. We consider the isotropic QHM Hamiltonian:
\begin{eqnarray}
\!\!\!\!\!\!H=-J_1\sum_{<ij>}\Hat{\mathbf{S}}_i\cdot\Hat{\mathbf{S}}_j-J_2\sum_{\ll ij\gg}\Hat{\mathbf{S}}_i\cdot\Hat{\mathbf{S}}_j-\sum_i \mathbf{B}_i\cdot\Hat{\mathbf{S}}_i, \label{Ham}
\end{eqnarray}
where $\hat{\mathbf{S}}_i$ is the (vector) spin operator associated with a 3-component spin at site $i$, 
$\mathbf{B}_i$ is the external magnetic field at site $i$, and $<ij>$ 
($\ll ij \gg$) denote, 
respectively, nearest-neighbor (NN) and next nearest-neighbor (NNN) sites, with  
coupling constants $J_1$ ($J_2$).

The double-time Green's function is defined as
\begin{eqnarray}
    i G_{mn}^{\alpha\delta}(t_1,t_2)\equiv \langle T[\Hat{S}_m^\alpha(t_1) \Hat{S}_n^\delta(t_2)]\rangle,
\end{eqnarray}
where $T$ is the usual time-ordering operator and $\langle\hspace{1em}\rangle$ denotes a ground state average at zero temperature. The superscripts in Greek letters refers to the spin components $x,y,z$, while $mn$ are site indexes, and for the time variables we henceforth use a simplified notation $t_i\rightarrow i$. With the spin ladder operators defined as $\hat{S}^{\pm}=\hat{S}^x\pm i\hat{S}^y$, the number of spin components can also be enlarged, i.e. $\alpha,\delta \in \{x,y,z\}$ or equivalently, $\alpha,\delta \in \{+,-,z\}$. 
 When $t_1-t_2$ is infinitesimal positive difference, the Green's function provides the ground state correlation function, i.e.
\begin{eqnarray} \label{Gdefin}
    i G_{mn}^{\alpha\delta}(1^+,1)=\langle \Hat{S}_m^\alpha \Hat{S}_n^\delta\rangle.
\end{eqnarray}
The property of the propagator from one space-time point to another is fully expressed with the components representing spin combinations $\lbrace xx, xy, xz,\cdots,zz \rbrace$, which can be written in matrix form, or equivalently, with the independent components $\lbrace +-, zz \rbrace$. For example, the equation of motion (eom)  for $G_{mn}^{+-}$ is 
\begin{eqnarray}
   && i\partial_{t_1}G_{mn}^{+-}(1,2)=\nonumber\\
   &&-2i\sum_{\ti\neq m}J_{m\ti}\Big[G^{(3)}{}_{m\ti n}^{z+-}(1,1,2)-G^{(3)}{}_{\ti mn}^{z+-}(1,1,2)\Big]\nonumber\\
   &&+B_m^z(1)G_{mn}^{+-}(1,2)+2\delta(1-2)\delta_{mn}\ex{\hat{S}_m^z(1)}, 
\label{eq:eom}    
\end{eqnarray}
where 
\begin{eqnarray}
    G^{(3)}{}^{\alpha\beta\delta}_{m\ti n}(1,1,2)\equiv \langle T[\Hat{S}_m^\alpha(1)\Hat{S}_{\ti}^\beta(1)\Hat{S}_n^\delta(2)]\rangle
\end{eqnarray}
is a higher-order GF, composed of three field operators,
and $\ti$ labels the sites with non-zero exchange coupling to site $m$. For our system,  $J_{m\ti}=J_{1}$ ($J_2$) when $\ti$ is the NN (NNN) site of $m$.

Tyablikov's method approximates the higher-order Green's function with
\begin{eqnarray}
    &G&^{(3)}{}^{\alpha\beta\delta}_{m\ti n}(1,1,2)\simeq \langle\Hat{S}_m^\alpha\rangle \langle T[\Hat{S}_{\ti}^\beta(1)\Hat{S}_n^\delta(2)]\rangle+\nonumber\\
    &&\langle\Hat{S}_{\ti}^\beta\rangle\langle T[\Hat{S}_m^\alpha(1)\Hat{S}_n^\delta(2)]\rangle,\label{G3T}
\end{eqnarray}
which works well with pure ferromagnetic or anti-ferromagnetic systems. However, for frustrated systems, the simple factorization in Eq.~(\ref{G3T}) leads to discrepancies. In the following, the functional derivative method is used to derive the self-consistent equation for the GF, giving a rigorous justification of Tyablikov's approximation, and extending the formulation to improve over it.

\begin{figure*}[t]
	\centering
	\includegraphics[scale=0.66]{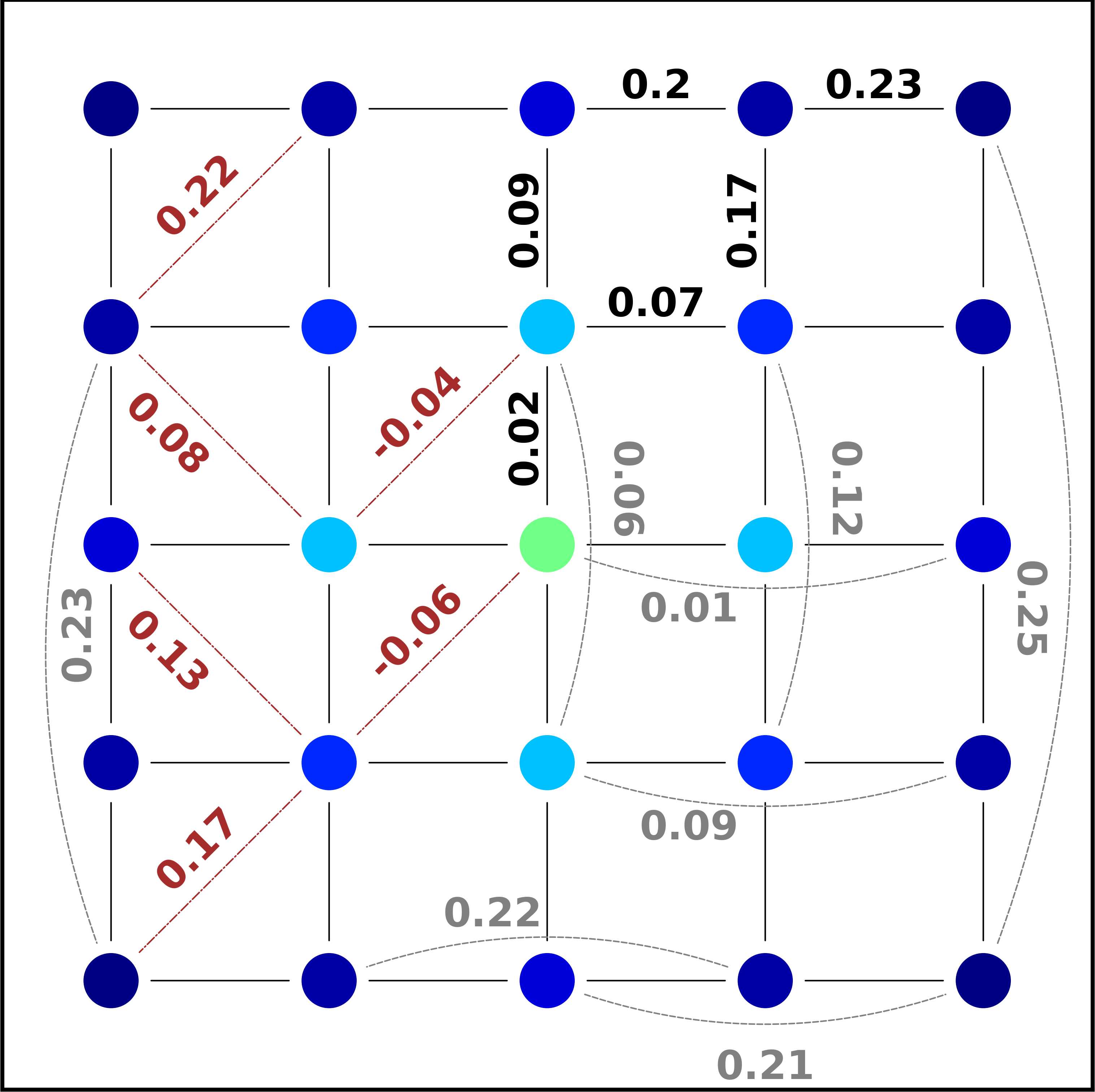}
	\includegraphics[scale=0.66]{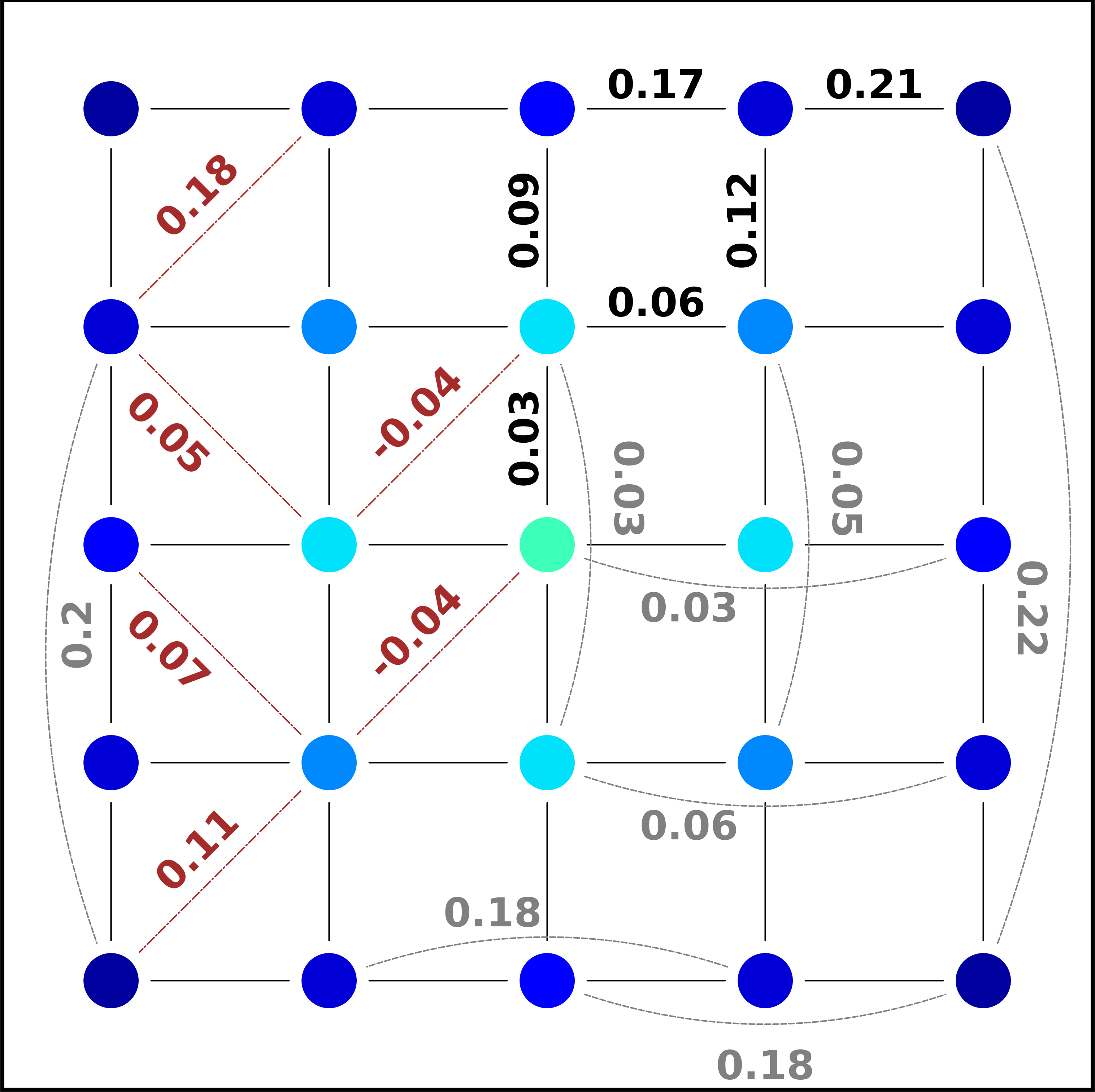}
    \includegraphics[scale=0.66]{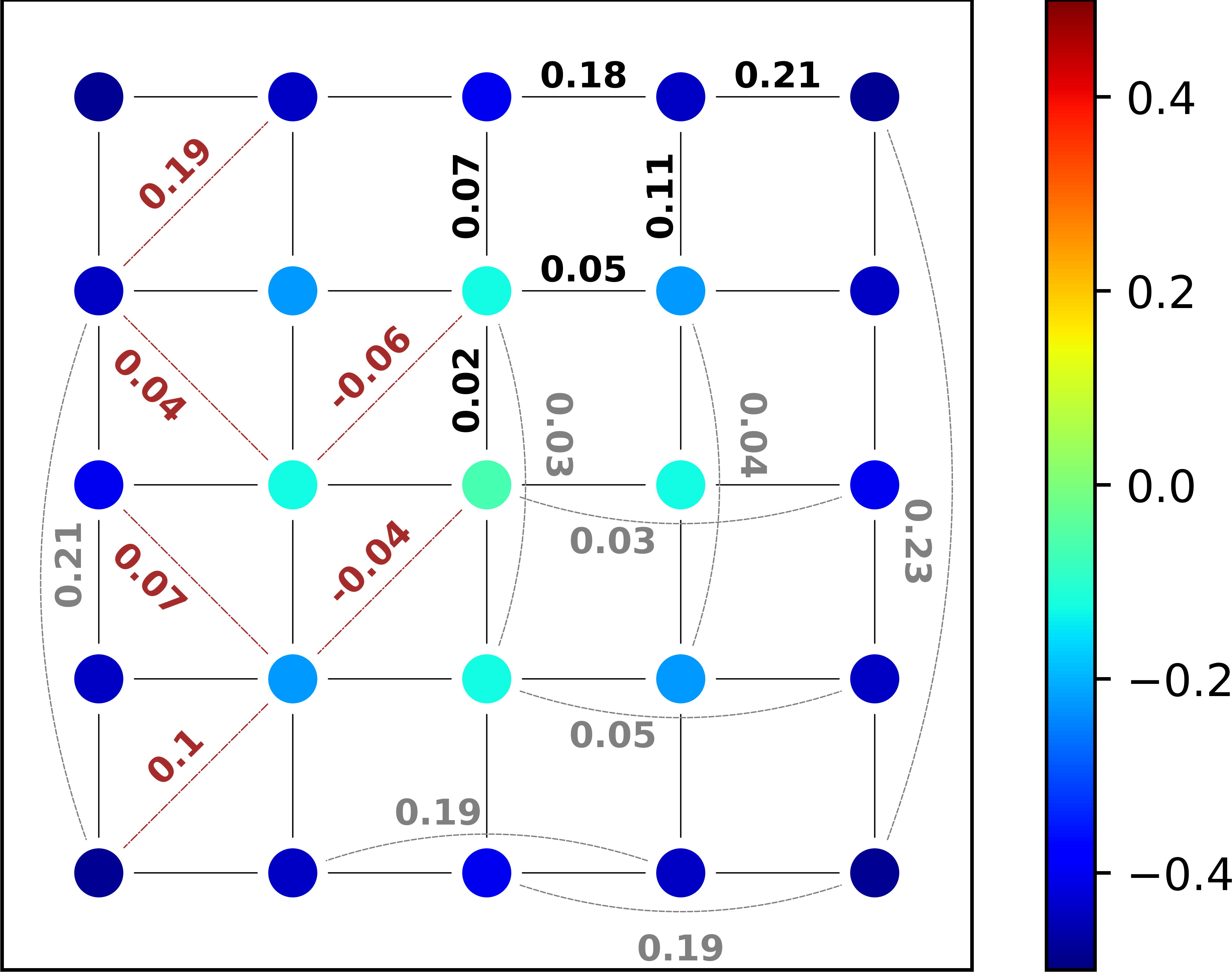} 
	\caption{Comparison between RPA (left panel), ED (middle panel) and GF (right panel)
	results for a $5\times 5$ lattice with open boundary conditions.
	The results are for the $S_\mathrm{total}^z = 17/2$ subspace, with FM exchange parameters: $J_1 = 1, J_2 =-0.5$. The color coding in the vertical bar applies to all panels.}
        \label{fig:vs}
\end{figure*}

\subsection{Schwinger derivative technique}
\label{sec:local}
In the interaction picture,
\begin{eqnarray}
    i G_{mn}^{\alpha\delta}(1,2)\equiv \frac{\langle\Psi| T[\hat{U}\Hat{S}_m^\alpha(1) \Hat{S}_n^\delta(2)]|\Psi\rangle}{\langle\Psi|\hat{U}|\Psi\rangle},
\end{eqnarray}
where a local probing field $\mathbf{B}$ is contained in the evolution operator,
\begin{eqnarray}
    \hat{U}=T \exp{\int_{-\infty}^{\infty}d1 \sum_i\Hat{\mathbf{B}}_i(1)\cdot\mathbf{S}_i(1)}.
\end{eqnarray}
The higher-order Green's function can be related to the response of the Green's function to the component of the local probing field:
\begin{eqnarray}
\!\!\!\!\! i\frac{\delta G_{mn}^{\alpha\delta}(1,2)}{\delta B_l^\beta(3)}=iG^{(3)}{}_{lmn}^{\beta\alpha\delta}(3,1,2)-G_{mn}^{\alpha\delta}(1,2)\langle \hat{S}_l^{\beta}(3)\rangle
\end{eqnarray}
A self-energy can now be defined making use of the eom and the functional derivative. Considering for concreteness 
the case $G_{mn}\equiv G_{mn}^{+-}$, 
\begin{eqnarray}
   && \sum_i\int d3\Sigma_{mi}(1,3)G_{in}(3,2)\equiv\nonumber\\
   &&2i\sum_{\ti}J_{m\ti}\Big(\frac{\delta G_{mn}(1,2)}{\delta B_{\ti}^z(1)}-\frac{\delta G_{\ti n}(1,2)}{\delta B_m^z(1)}\Big).
\end{eqnarray}
An Hartree-like potential and an exchange-like potential can also be defined:
\begin{eqnarray}
    V_m^H(1)&=\sum_{\ti}J_{\ti m}\ex{\hat{S}_m^z(1)}\\
    V_{m\ti}^F(1)&=J_{\ti m}\ex{\hat{S}_m^z(1)},
\end{eqnarray}
leading to a Dyson-like eom in terms of the self-energy:
\begin{eqnarray}
[i\partial_{t_1}-V_m^H(1)-B_m^z(1)]G_{mn}(1,2)-\sum_{\ti\neq m}V_{m\ti}^FG_{\ti n}(1,2)\nonumber\\
\!\!\!\!\!\!\!-\sum_i\int d3\Sigma_{mi}(1,3)G_{in}(3,2)=2\delta(1-2)\delta_{mn}\ex{\hat{S}_m^z(1)}.\nonumber\\
\end{eqnarray}
Using the identity $ \frac{\delta G}{\delta B}G^{-1}+G\frac{\delta G^{-1}}{\delta B}=0$,
the self-energy can be re-cast as
\begin{eqnarray}
    &&\Sigma_{mn}(1,2)=-i\sum_{\ti\neq m,l}J_{m\ti}\int d3\times\nonumber\\
   && \Big\lbrace G_{ml}(1,3)\frac{\delta G_{ln}^{-1}(3,2)}{\delta B_{\ti}^z(1)}-G_{\ti l}(1,3)\frac{\delta G_{ln}^{-1}(3,2)}{\delta B_m^z(1)}\Big\rbrace.
\end{eqnarray}
To solve for the response of $G$ with respective to $B$, we take functional derivative of Eq.~\eqref{eq:eom} and neglect the second derivative term $\delta^2 G/\delta B^2$, thus obtaining
\begin{eqnarray}
    &&i\partial_1\frac{\delta G_{mn}(1,2)}{\delta B_l^z(3)}= G_{mn}(1,2)\delta(1-3)\delta_{ml}+\nonumber\\
    &&\sum_{\ti\neq m}J_{m\ti}\bigg\lbrace\frac{\delta \ex{\hat{S}_{\ti}^z(1)}}{\delta B_l^z(3)}G_{mn}(1,2)+\ex{\hat{S}_{\ti}^z(1)}\frac{\delta G_{mn}(1,2)}{\delta B_l^z(2)}-\nonumber\\
    &&\frac{\delta \ex{\hat{S}_m^z(1)}}{\delta B_l^z(3)}G_{\ti n}(1,2)-\ex{\hat{S}_m^z(1)}\frac{\delta G_{\ti n}(1,2)}{\delta B_l^z(2)}\bigg\rbrace+\nonumber\\
    &&2\frac{\delta \ex{\hat{S}_m^z(1)}}{\delta B_l^z(3)}\delta_{mn}\delta(1-2).
\end{eqnarray}
To solve for the response function
    $R_{ml}(1,2)=\frac{\delta\ex{S^z_m(1)}}{\delta B_l^z(2)}$
we look at the eom for $\ex{\hat{S}_m^z(1)}$:
\begin{eqnarray}
    &&\partial_1\frac{\delta \ex{\hat{S}_m^z(1)}}{\delta B_l^z(2)}=i\sum_{\ti\neq m}J_{m\ti}\frac{\delta}{\delta B_l^z(2)}\ex{\hat{S}_i^+(1)\hat{S}_m^-(1)}\nonumber\\
    &&=\sum_{\ti\neq m,pq}J_{m\ti}\int d3d4G_{\ti p}(1^+,3)\frac{\delta G_{pq}^{-1}(3,4)}{\delta B_l^z(2)}G_{qm}(4,1)\nonumber\\
\end{eqnarray}
Starting with $G$ computed at the mean-field level, the set of equations (13), (15)-(17) can be solved self-consistently. Then, using 
Eq.~\eqref{Gdefin}, the observables as ground state expectation values and correlation functions can be computed from the equal-time Green's function.
\subsection{Solving the equations in momentum space}
The magnon Hedin's equations in the last section were expressed in real space. Considering the time translational and spatial symmetries of the system, a Fourier transform provides the GF in momentum-frequency space:
\begin{align}
    G_{mn}(1,2)  &\equiv G_{mn}(t_1,t_2)\nonumber\\
    &= \int d\omega  \int d\mathbf{k} G(\mathbf{k},\omega)e^{i\mathbf{k}\cdot(\mathbf{r}_m-\mathbf{r}_n)} e^{-i\omega(t_1-t_2)}.
\end{align}
The vertex function is now defined as 
\begin{align}
    \Lambda_{p,q,l}(1,2,3)&\equiv\frac{\delta G_{pq}^{-1}(1,2)}{\delta B^z_l(3)},
\end{align}
which can be written in momentum-frequency space accordingly:
\begin{align}
    \Lambda_{p,q,l}(1,2,3)&\equiv \Lambda_{p,q,l}(t_1,t_2,t_3)\nonumber\\
    &=\int d\mathbf{k}d\mathbf{k}'d\omega d\omega' e^{i\mathbf{k}\cdot(\mathbf{r}_p-\mathbf{r}_q)}e^{i\mathbf{k}'\cdot(\mathbf{r}_p-\mathbf{r}_l)}\nonumber\\
    &\times e^{-i\omega(t_1-t_2)}e^{-i\omega'(t_1-t_3)}\Lambda(\mathbf{k},\mathbf{k}';\omega,\omega').
\end{align}
In the $(\mathbf{k},\omega)$ space, our equations read
\begin{eqnarray}
&&\!\!\!\!\!\!\big[\omega-V^\mathrm{HF}-B^z\big] G(\mathbf{k},\omega) = \Sigma(\mathbf{k},\omega)G(\mathbf{k},\omega)+\ex{S^z}\\
&&\!\!\!\!\!\!\omega R(\mathbf{k},\omega)=J(\mathbf{k})\int d\mathbf{k}'d\omega'\nonumber\\
&&\!\!\!\!\!\!\times~G(\mathbf{k}+\mathbf{k}';\omega+\omega')\Lambda(\mathbf{k},\mathbf{k}';\omega,\omega') G(\mathbf{k}',\omega').
\end{eqnarray}
Here, $V^\mathrm{HF}=V^\mathrm{H}+V^\mathrm{F}$, and $V^\mathrm{H}, V^\mathrm{F},\ex{S^z}$ are the Fourier transform of their corresponding real space-time values.

\section{Results and discussions}
In this Section, the approach introduced in Sect.~\ref{theorysect}
is applied to 2D Heisenberg systems with square and hexagonal lattices, and different types of exchange coupling. To assess the performance of the method, the Green's function (GF) results are compared with numerical benchmarks from the the Exact Diagonalization (ED) method.
\subsection{The case of a 2D square-lattice cluster}
The system we consider is a $5\times5$ lattice with open boundary conditions.
We discuss both FM and AFM magnetic regimes in
few selected subspaces with  total spin projection $S_\mathrm{total}^z$. Compared to either the FM limit $(25\uparrow, 0\downarrow)$ or the AFM one $(13\uparrow,12\downarrow)$,
an intermediate value of $S_\mathrm{total}^z$ shows most clearly the competition of NN and NNN exchange couplings. Thus it is convenient to start the discussion with the subspace 
$S_\mathrm{total}^z=17/2$  $(21\!\uparrow,4\!\downarrow)$. We use the parameters $J_1=1,J_2=-0.5$. The results are shown in Fig.~\ref{fig:vs}, where we compare RPA, ED and GF results. The
color palette is used to represent the expectation value of the $z-$component spin $\ex{S_m^z}$, and the numbers show the $z$-$z$ spin correlation between lattice points $\ex{S_m^zS_{n}^z}$. 

Compared to the RPA decoupling method, 
the inclusion of the self-energy in the Green's function method
gives higher accuracy for both $\ex{S_m^z}$ and  $\ex{S_m^zS_{\bar{m}}^z}$, 
as shown by the improved locations of the poles of the GF.
The reason behind the improvement is that the direct response $\delta G/\delta B$, which is either treated as a constant (possibly with value 0) in the RPA method, gives a nonzero dynamical contribution to the self-energy. Therefore, the approximation used in the derivation of self-consistent equations in section \ref{sec:local} is the key step forward compared to the bare decoupling. Including higher-order responses $\delta^n G/\delta B^n$ can in principle improve the accuracy, but at the cost of increased converging difficulties and heavier computational burden.

\begin{figure}[t]
\centering
\includegraphics[scale=0.4]{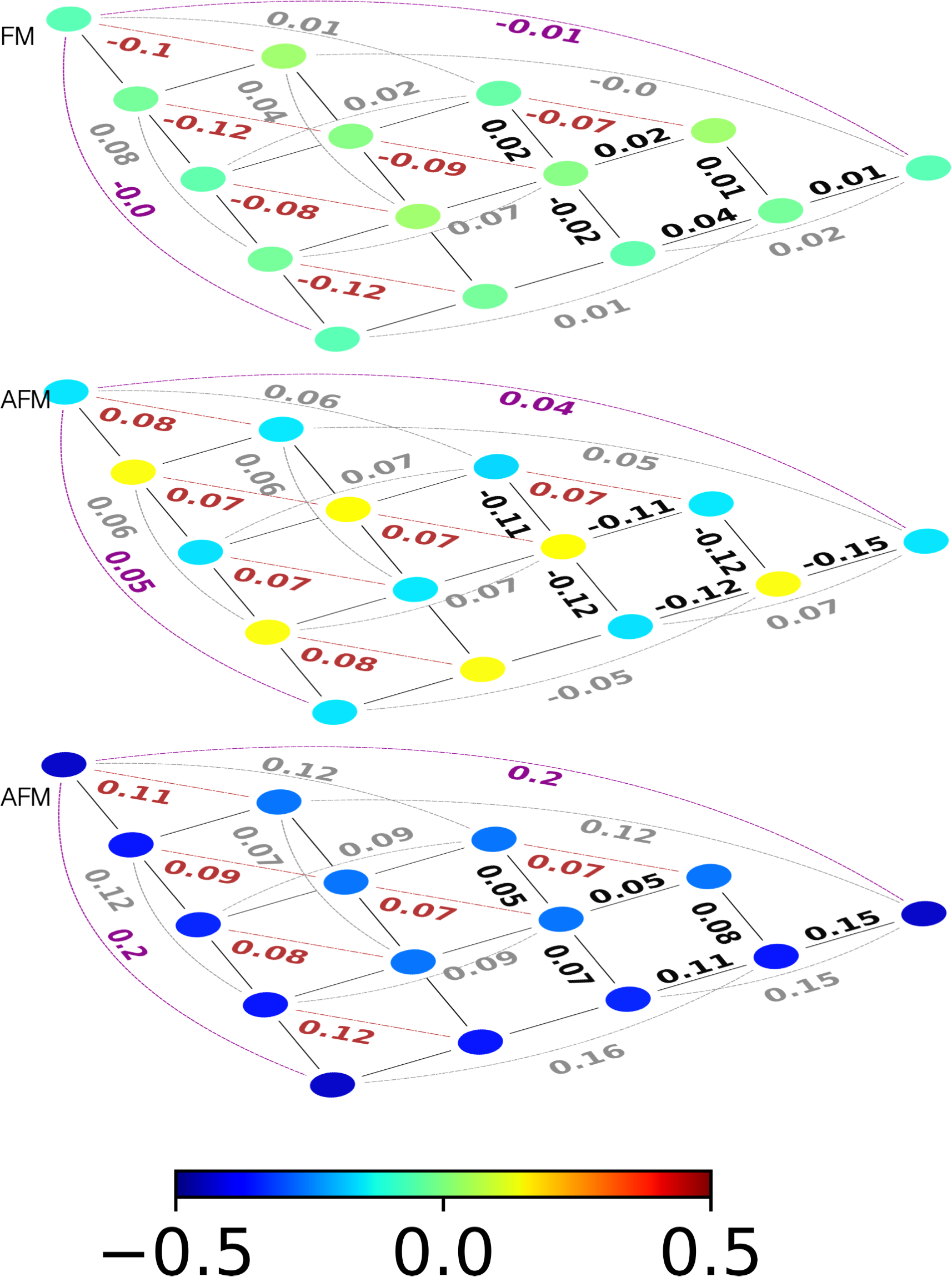}\\
%
%
\caption{$\langle S_m^z\rangle$ (denoted by color) and $\langle S_m^zS_{n}^z\rangle$ (denoted by numbers) for a $5\times5$ square lattice system with open boundary conditions. The color coding in the vertical bar applies to all cases, and results in each panel fulfill the C$_{4v}$ square symmetry. Bottom panel: AFM, $S_\mathrm{total}^z=17/2$; Middle panel: AFM, $S_\mathrm{total}^z=1/2$; Top panel: FM, $S_\mathrm{total}^z=1/2$. 
The AFM coupling parameters are $J_1=-1,J_2=0.5$ and the FM ones are $J_1=1,J_2=-0.5$.}
\label{fig:sz} 
\end{figure}
 
Our self-energy GF shows good accuracy also for $J_1=-1,J_2=0.5$, $S_\mathrm{total}^z=17/2$. Such parameters lead to pure AFM interaction (i.e. no frustration) on the square lattice. In the $S_\mathrm{total}^z=17/2$ subspace, where the majority of the configurations is with spin up, the ground state  due to the AFM couplings is relatively homogeneous. This is detailed  in Fig.~\ref{fig:sz}, where 
$-0.45\leq\langle S_m^z\rangle\leq-0.27$ and $\langle S_m^zS_{n}^z\rangle>0$ for all lattice sites.

With the same couplings, but for the $S_\mathrm{total}^z=1/2$ subspace,  the GF method describes well the Neel-type ground state: the distribution of $\langle S_m^z\rangle$ is bipartite, sites on the same/different sublattices are positively/negatively correlated. 

Remaining in the $S_\mathrm{total}^z=1/2$ subspace, but this time with $J_1=1,J_2=-0.5$, we observe that the GF ground state has a small total spin value: the magnitudes of $\langle S_m^z\rangle$ are close to zero, NN sites are weakly correlated compared with 
$S_\mathrm{total}^z=1/2, J_1=-1,J_2=0.5$. Such behavior is reminiscent of what occurs for systems with an even number of sites, where Lieb's theorem states  that $S_\mathrm{total}=0$ in the ground  state.  For the three scenarios discussed above, the ground states are either relatively homogeneous (the signs of the exchange couplings and the
net value of  $S_\mathrm{total}^z$ in the given subspace are chosen so that they impose conflicting constraints on
the spin alignment) or bipartite.
This suggests that, for these cases, quantum fluctuations introduced by higher-order response terms play only a 
small role in the determination of the Green's function.

\begin{figure}[t]
    \includegraphics[scale=0.09]{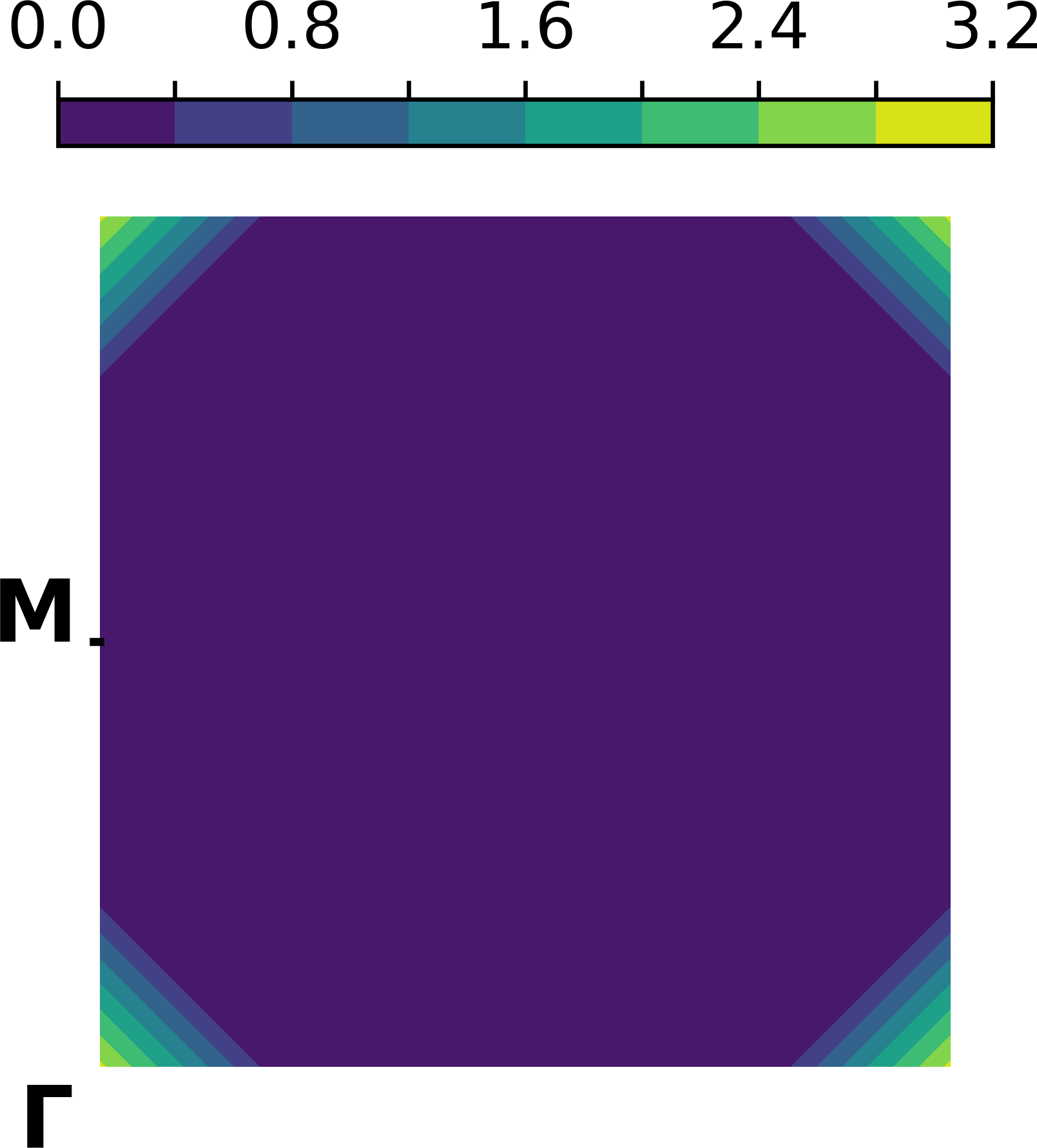}\hspace{0.5em}
    \includegraphics[scale=0.09]{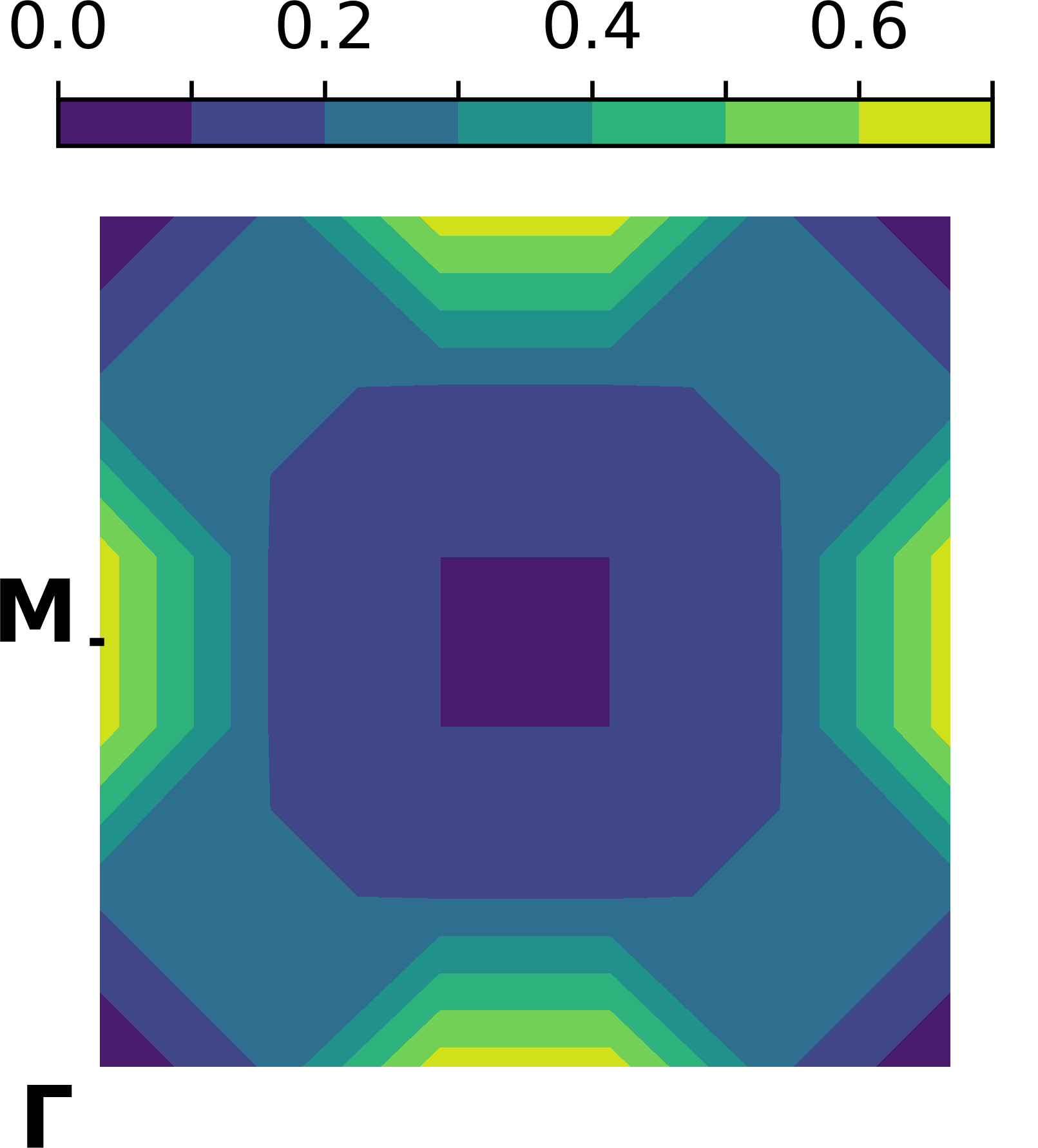}
    \caption{Static structure factor $S(\mathbf{q})$ ($\mathbf{q}\in[0,2\pi]\times[0,2\pi]$) of a $5\times5$ square lattice with periodic boundary conditions and $J_1=1,J_2=-0.5$.
    Left: $S_\mathrm{total}^z=17/2$;
    Right: $S_\mathrm{total}^z=1/2$. The high symmetry points in the first Brillouin Zone are labeled by $\mathbf{\Gamma}\equiv (0,0)$, and $\mathbf{M} \equiv (0,\pi),(\pi,0)$.}
    \label{fig:sq}
\end{figure}

Additional perspective on the method performance can be gained from the spin structure factor, 
defined as 
\begin{align}
 S(\mathbf{q})=\frac{1}{N^2}\sum_{mn}\ex{S_m^zS_n^z}e^{i\mathbf{q}\cdot(\mathbf{R}_m-\mathbf{R}_n)},
 \end{align}
for a $5\times5$ lattice with periodic boundary condition. Because we are considering a square (i.e. bipartite) lattice, a Neel-like order is not compatible
with a cluster with an odd number of atoms and periodic boundary conditions. Thus, we consider a 
 $5\times5$ ($4\times4$) cluster for FM (AFM) NN exchange. The GF results for $S(\mathbf{q})$ for a $5\times 5$ FM  cluster are shown in Fig.~\ref{fig:sq}. The structure factor is strongly peaked at $\mathbf{\Gamma}$ point for $S^z_\mathrm{total}=17/2$ and relatively weakly peaked at $\mathbf{M}$ points for $S^z_\mathrm{total}=1/2$, which agrees with 
previous results in the literature \citep{shannon2004finite}.
The displayed GF results are in very good agreement with the ED ones, and
the differences between the two treatments are indistinguishable on the scale of the figure. Similar consideration apply to the agreement between GF and ED methods
for a  $4\times4$ cluster (not shown).
 
\subsection{Single- and double-impurity configurations}
\begin{figure}[t!]
     \centering
     \includegraphics[scale=0.4]{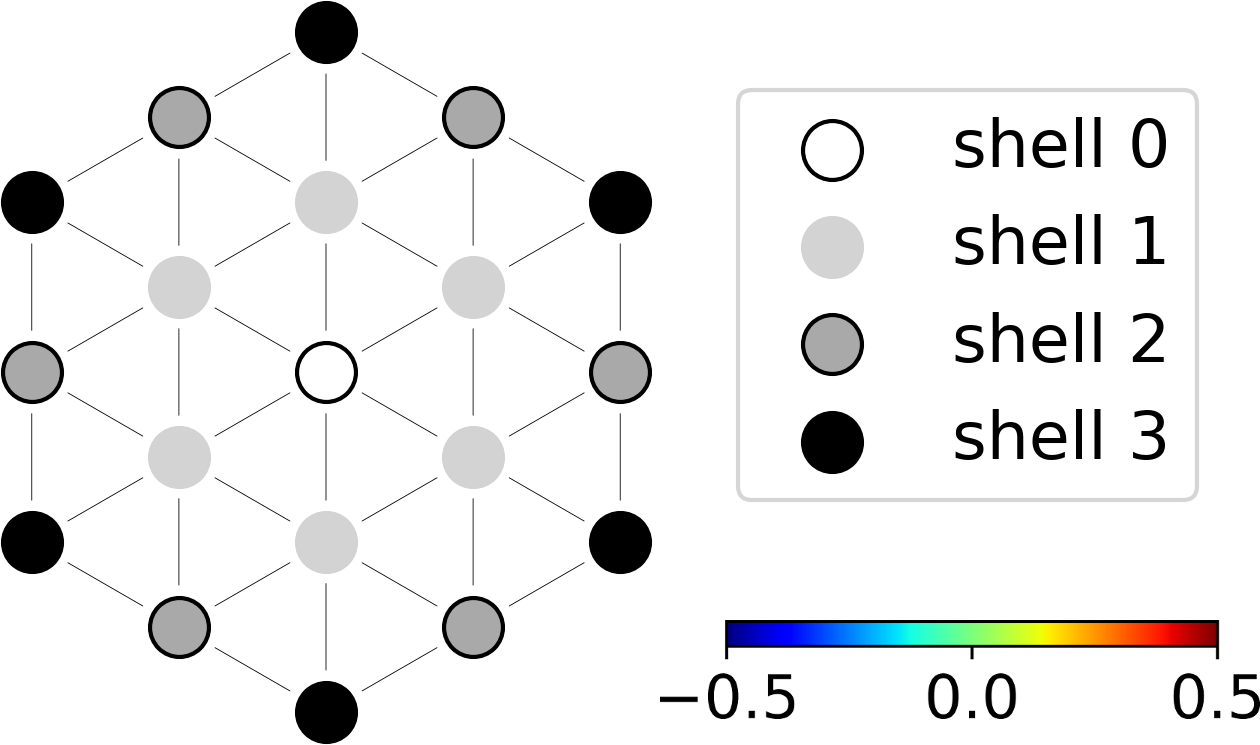}\\\vspace{1em}
         \includegraphics[scale=0.6]{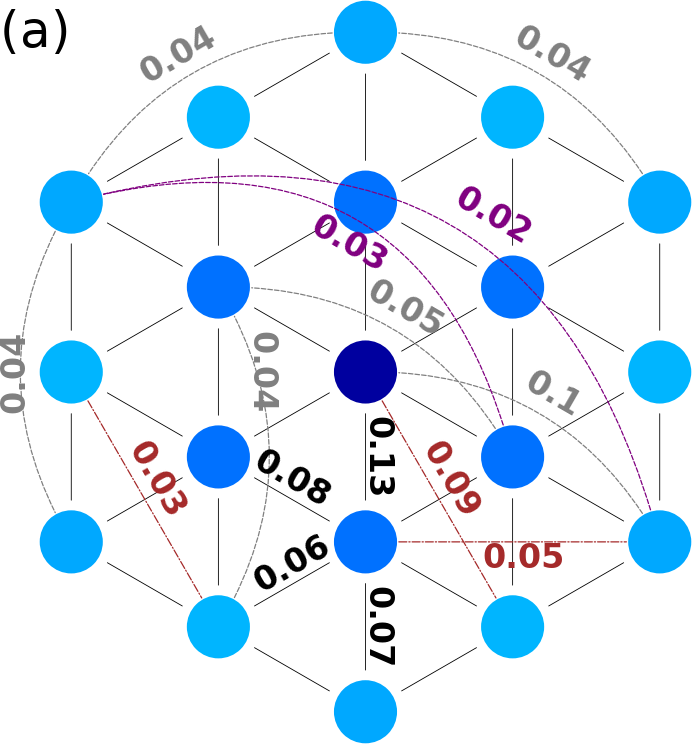}\hspace{1em}
         \includegraphics[scale=0.6]{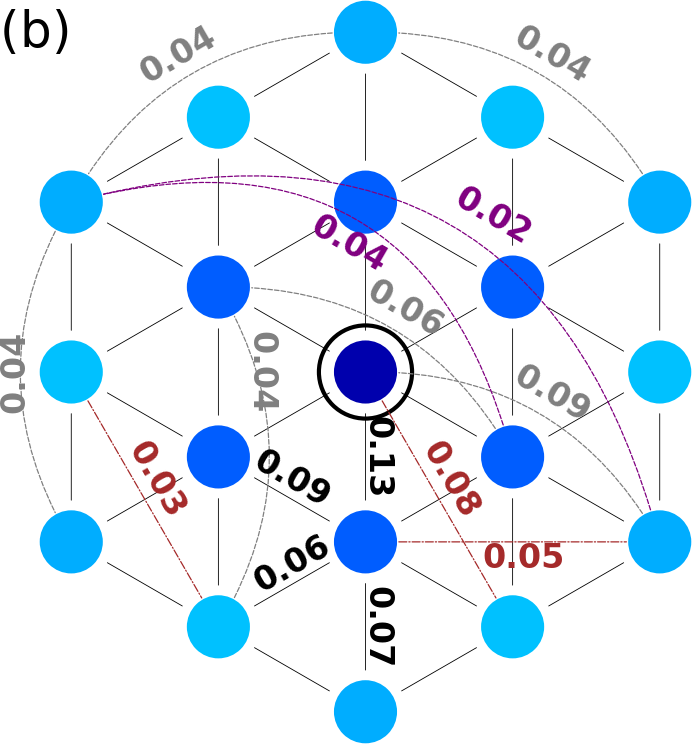}\vspace{1em}
         \includegraphics[scale=0.6]{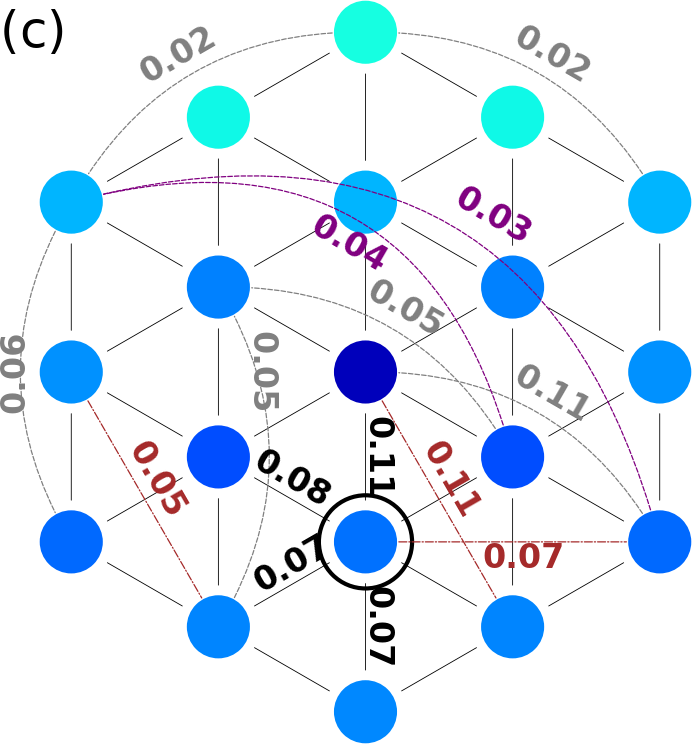}\hspace{1em}
         \includegraphics[scale=0.6]{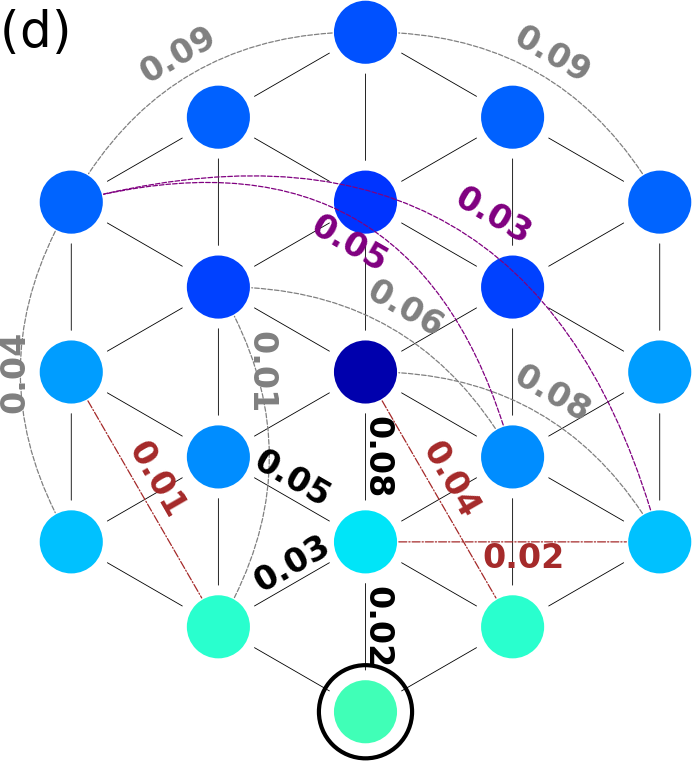}\vspace{1em}
         \includegraphics[scale=0.6]{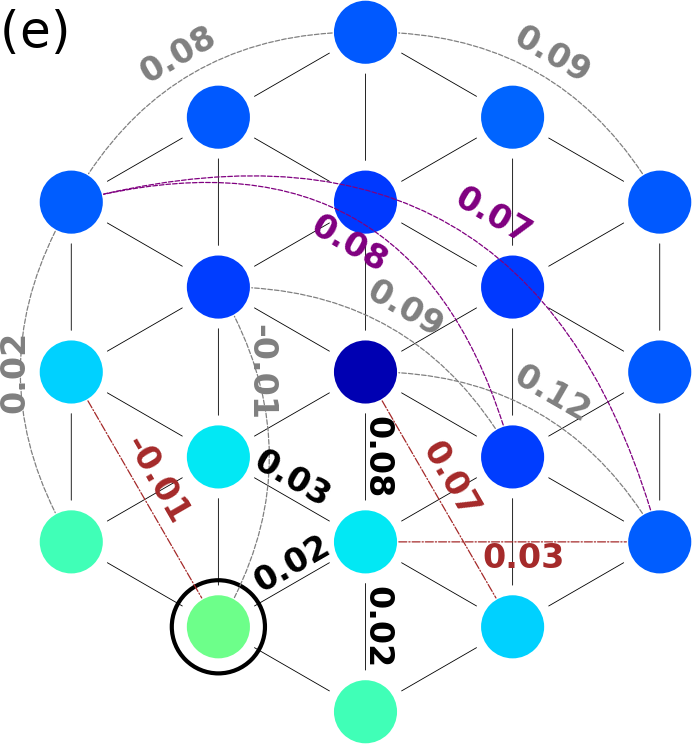}\hspace{1em}
         \includegraphics[scale=0.6]{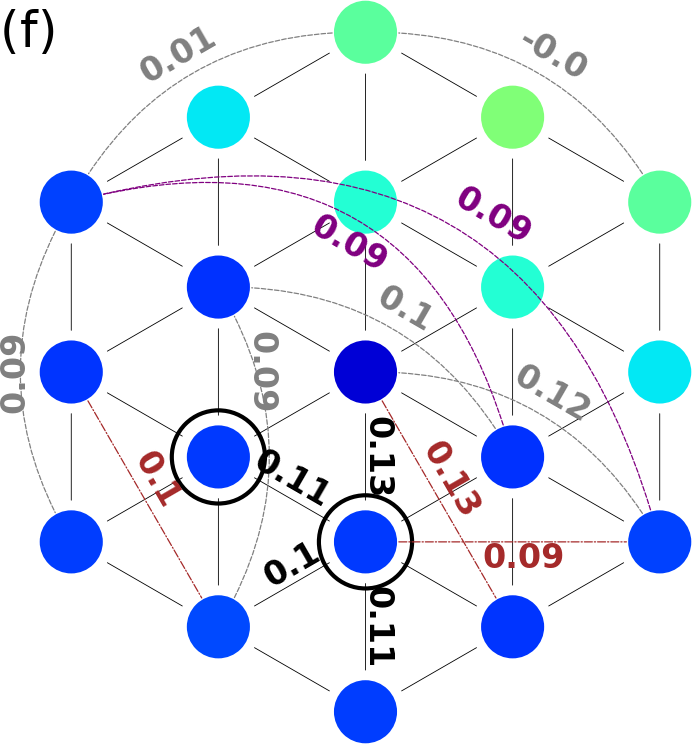}
     \caption{ $\langle S_m^z\rangle$ (denoted by color) and $\langle S_m^zS_{n}^z\rangle$ (denoted by  numbers) in an open-boundary 19-site hexagonal lattice system with and without impurity. The non-impurity coupling parameters are  $J_1=1,J_2=-0.5$. The additional impurity coupling strengths are $\Delta J_{NN}=0.5J_1,\Delta J_{NNN}=0.5J_2$. $S^z_\mathrm{total}=9/2$ $ (14\uparrow5\downarrow)$. The top panel illustrates different shells of atoms (see main text). The color coding in the horizontal bar applies to all cases. (a) No impurity case. (b)-(f) The circles with a
     black outline are impurity sites.}
     \label{fig:hex}
\end{figure}

In realistic cases, one is often faced with the problem of having impurities in
the system under investigation.
Impurity atoms can be included by introducing an additional term in the Hamiltonian,
\begin{eqnarray}
    H_\mathrm{imp}=-\sum_{ij}\Delta J_{ij}\Hat{\mathbf{S}}_i\cdot\Hat{\mathbf{S}}_{j}.
\end{eqnarray}
where either $i$ or $j$ denote the impurity site(s).
In the following, we specialize to the cases of single and double impurities in a 19-site hexagonal lattice
and choose to work in the subspace $S^z_\mathrm{total}=9/2$,which, as for the square lattice, nicely
illustrate the interplay of FM and AFM couplings.
In the no-impurity case, the coupling parameters
are $J_1=1,J_2=-0.5$; in the presence of impurities, we have the additional coupling strengths $\Delta J_{ij,NN}=0.5J_1,\Delta J_{ij,NNN}=0.5J_2$. 
The GF result  is shown in Fig.~\ref{fig:hex}. 
It is convenient for the discussion to organize the lattice sites in shells, where sites in a given shell are
equally distant from central site, and different shells correspond to different distances (Fig.~\ref{fig:hex}).

For the non-impurity case (Fig.~\ref{fig:hex}a), and because of the FM NN couplings and the $C_{6v}$ lattice symmetry,  the central spin assumes the spin-down $\downarrow$ configuration. 
With $\ex{S_m^z}<0$ and $\ex{S_m^zS_n^z}>0$ at all sites, we then conclude that the non-impurity system is dominated by FM interactions. 

Inserting one impurity in the system amplifies both NN and NNN couplings. Locating the impurity at the center (Fig.~\ref{fig:hex}b) effectively increases the FM strength around the impurity, which can be seen from the increased correlation between the impurity site and its NN.
When the impurity moves away from the cluster center, the $C_{6v}$ symmetry is broken. 
If the impurity is in shell 1 (Fig.~\ref{fig:hex}c), the number of its FM NN sites remains 6, while the number of its AFM NNN sites decreases. Accordingly, the couplings between the impurity and its NN are FM dominated, and thus
the spins maintain the $\downarrow$ configuration.
However, when the impurity atom moves to the boundary of the lattice (Fig.~\ref{fig:hex}d,e), 
the value of spin-$z$ projection at the impurity, $\ex{S_I^z}$,  is close to zero. 
This change of $\ex{S_I^z}$ when moving from the center towards the cluster boundary 
(where the number of NN and NNN sites is smaller),
can be ascribed to the change in the number of neighbors, i.e. finite size effect and  the 
geometry of the cluster play an important role. 
Finally, we also show results for one geometry with two impurities,
where the latter are both located in shell 1 and NN to each other (Fig.~\ref{fig:hex}f). In this case, 
the impurities and their NN spins are strongly FM coupled, and form a  small FM sub-cluster. 

The results presented in Fig.~\ref{fig:hex}, which were obtained with our novel GF technique,
compare very well with ED calculations (not shown). Furthermore, the trends for other $S^z_\mathrm{total}$ subspaces are
very similar, with the same level of agreement between GF and ED schemes. 

As an overall remark to this section,
the GF approach appears to be able to capture all the effects due to 
the $J_1-J_2$ competition, also in the presence of significant finite size effects. However,
it should also be noted that the type of spin-spin interactions considered in the paper  so far
are symmetric in nature (i.e, expressed in terms of scalar products between spins). In many
materials, the spin-orbit interaction can mediate anti-symmetric exchange couplings 
among spins. This more challenging situation is addressed in the next section.

\subsection{Including Dzyaloshinskii-Moriya interactions}
The Dzyaloshinskii-Moriya interaction (DMI), also referred to as antisymmetric exchange \citep{dzyaloshinsky1958thermodynamic,moriya1960anisotropic}, results from the interplay of spin-orbit and super-exchange interactions. The DMI can be written as  
\begin{align}
    H_\mathrm{DM} = D\sum_{<ij>}\mathbf{e}_{ij}\cdot(\hat{\mathbf{S}}_i\times\hat{\mathbf{S}}_j)
\label{DMIeq}
\end{align}
where $\mathbf{e}_{ij}=(\mathbf{r}_j-\mathbf{r}_i)/|\mathbf{r}_j-\mathbf{r}_i|$ is the unit vector pointing from site $i$ to site $j$. 
The DMI favours chiral canting of the spins, and is thus responsible for the emergence
of complex spin patterns, for example magnetic skyrmions.
Here, we wish to see how the GF method performs when DMI is present. To this end, we consider a system described by the Hamiltonian of Eq.~(\ref{Ham}) to which the
term $H_\mathrm{DM}$ of Eq.~(\ref{DMIeq}) is added. In the presence of DMI, a diagonalization partitioned in subspaces with definite $S_\mathrm{total}^z$ is not possible, and
thus we consider a rather small  ($3\times3$) isolated cluster subject to an external field $(0,0,B)$, with FM parameters  $J_1=1,J_2=0,B=0.1$. The value chosen for the DMI is $D=4$,  that for the cluster considered provides noticeable canting of the spins. 

\begin{figure}[ht]
     \centering
         \includegraphics[width=0.5\textwidth]{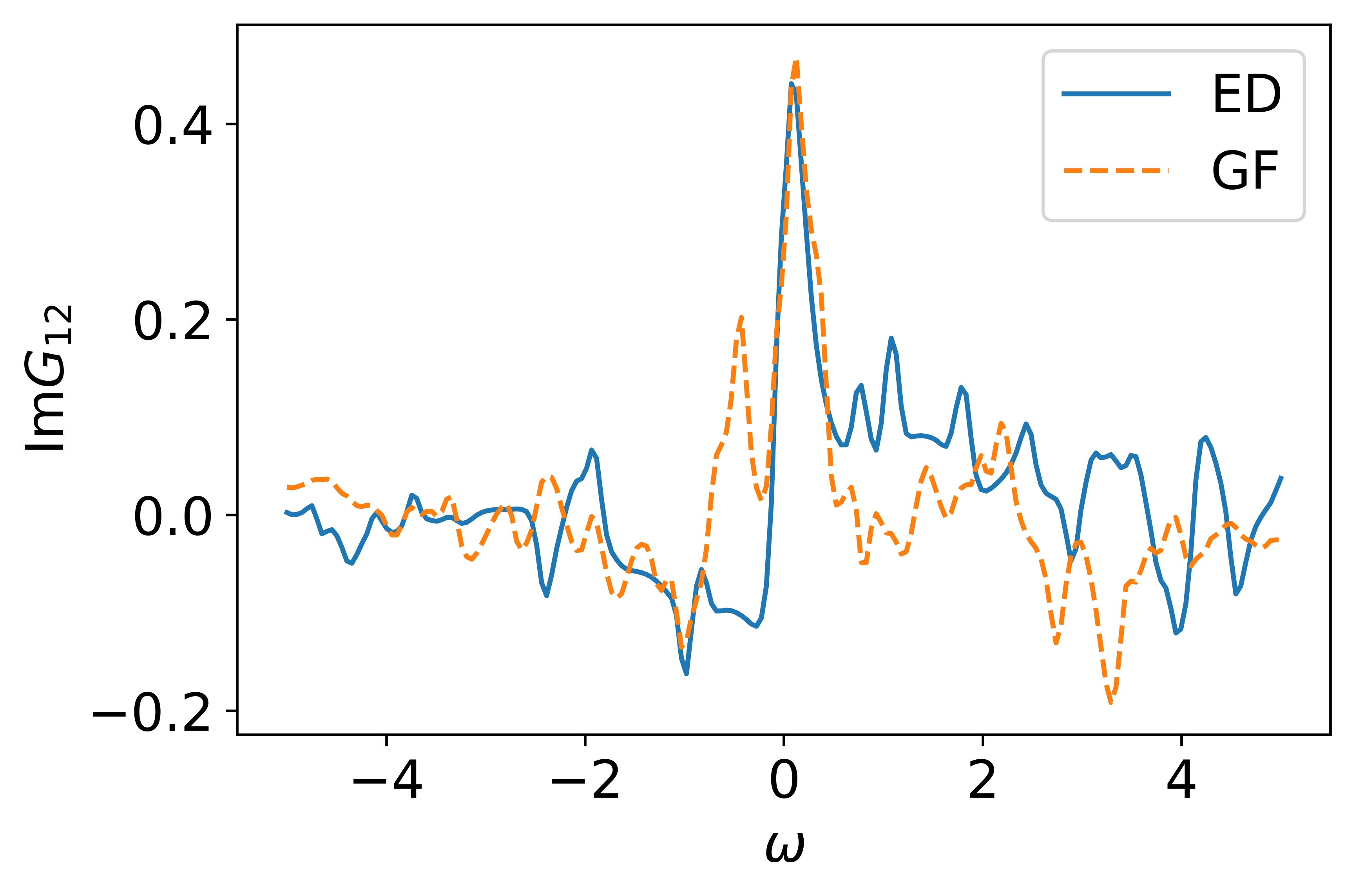}
         \caption{Imaginary part of the $G^{+-}_{12}$ component of the Green's function in the frequency domain.}
     \label{fig:ImG}
\end{figure}

\begin{table}[b]
\caption{Comparison between ED and GF on a $3\times3$ cluster. $a$ refers to the bottom-left site, $b$ refers to the left-middle site, and $c$ refers to the center site. The lattice symmetry requires that $\ex{S_c^x}=\ex{S_c^y}$=0, so neither of them is listed in the table.}
\label{tab:dmi}
    \begin{tabular}{c|c|c|c|c|c|c|c}
        & \hspace{0.5em}$\ex{ S_a^x }$\hspace{0.5em} & \hspace{0.5em}$\ex{S_a^y}$\hspace{0.5em} & \hspace{0.5em}$\ex{S_a^z}$\hspace{0.5em} & \hspace{1em}$\ex{S_b^x}$\hspace{0.5em} & \hspace{0.5em}$\ex{S_b^y}$\hspace{0.5em} & \hspace{0.5em}$\ex{S_b^z}$\hspace{0.5em} &\hspace{0.5em}$\ex{S_c^z}$\hspace{0.5em} \\
        \hline
        \hline
        ED & 0.05 & -0.05 & 0 & 0 &-0.13 & 0.05 & 0.27\\
        \hline
        \vspace{1mm}
        GF & 0.12 & -0.12 & 0.04 & 0 & -0.23 & 0.14 & 0.34\\
    \end{tabular}
\end{table}
On inclusion of the DMI, there is a significant discrepancy between GF and ED spin patterns,  (as is shown in Table~\ref{tab:dmi}, our GF approach fails to capture that the spins on the corner site of the cluster are completely in plane.), indicating that, at least for this small system, a large vertex correction beyond the RPA is needed 
(this cannot be easily provided by a low-order self-energy as 
the one considered in Sect.~\ref{theorysect}). To investigate the source
of such discrepancy, we consider the imaginary part of $G^{+-}_{12}$. In Fig.~\ref{fig:ImG} we show the results for $\Im G^{+-}_{12}$,
obtained both via our self-energy treatment and by applying spin operators on the ED ground state. 

The GF and ED curves show similar generic trends,
but for more detailed features the two treatments are clearly at variance, with
some specific structures differing both in position and 
strength. In particular, the locations of poles of $\text{Im}G$ from ED are overall more compactly distributed, especially at positive frequencies. 
This discrepancy, observed also for other GF components
(and for the real part of the GF as well) when DMI is present, is likely to be a general shortcoming of the GF approach within the linear response approximation and indicates the need
to include nonlinear response.
The quality of the GF solution in the presence of DMI is considerably better when we place the $3\times 3$ quantum spin cluster as  center block of a $9\times 9$ lattice,
and the remaining lattice points are occupied by mutually interacting classical spins with magnitude $|\Vec{S}|=1/2$, that also interact with the quantum spins.  Mixed quantum-classical systems can be
useful to gain insight in large systems when the spin pattern develops over 
several lattice distances, and a quantum treatment of all spins is computationally
not viable (as for example in the case of magnetic skyrmions \citep{gauyacq2019model}). 
For our $9\times 9$ system, we use a mixed quantum-classical self-consistent description where the quantum spins on the cluster border interact with the neigbhoring classical spins according to -$\sum_{qc}J_{qc}\hat{S}_q\cdot \vec{S}_c$.
\begin{figure}[ht]
    \centering
    \includegraphics[width=0.4\textwidth]{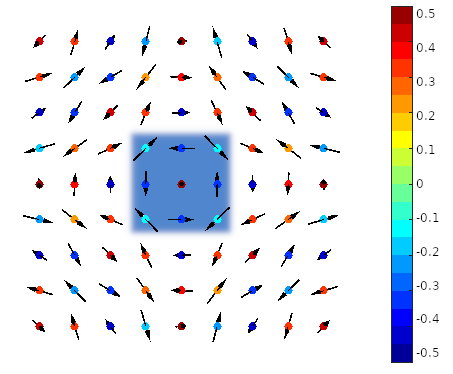}
    \caption{Quantum spins (shaded region) in the background of classical spins. The arrow represents the spin projection in the $xy$ plane, the color represents the $S^z$ for classical spins and $\ex{S^z}$ for quantum spins. The coupling parameters are the same as in the $3\times3$ pure quantum case.}
    \label{fig:QinC}
\end{figure}
For a suitable choice of the $J_1,J_2,D$ and $B$ parameters, 
a skyrmion-like ground state is expected to occur in the system. 
Here, we do not perform an extensive parameter search to establish
the skyrmion regime. Rather, and for sake of comparison,
the coupling parameters used are the same as in the isolated $3\times3$ 
quantum-spin cluster, and the results thus obtained are shown 
in Fig.~\ref{fig:QinC}. 
Within the figure resolution, the results from the self-energy approach and ED
are not distinguishable. Furthermore, they differ from those of the isolated 
cluster. From the figure, no skyrmion like patterns is easily discernible: however,
an interesting qualitative aspect is that ED and GF solution compare well, due to the 
action of the ``forcing" field due to the classical spins. 

Naturally, there is at this point no way to say for certain if  the similar ED and GF solutions describe well the full quantum exact solution for the $9 \times 9$ cluster. However,
as an overall and final remark to the results of this section we note that, 
unlike
the Heisenberg exchange coupling, the DMI involves the cross-product of the spin operators. Such difference in structure and symmetry of the Hamiltonian leads to a very different structure of the magnon Hedin's equations. In the derivation in section \ref{sec:local}, 
the higher-order response of the Green's function is approximated with zero. The approach shows good result for scalar-product exchange couplings, but probably requires more considerations 
for interactions involving cross-product couplings.

\section{Conclusion and outlook}

Remarkable progress has been made in understanding magnetism in condensed matter from a microscopic perspective. However, as of today, describing complex magnetic configurations in real materials largely remains an open problem. This is because long-ranged magnetic patterns emerge from a delicate balance of several factors: 
among the most important are electronic correlations, electron-phonon interactions, crystal field effects, spin-orbit interactions, disorder and impurities. Additionally, the number of atoms involved in the periodic unit (for commensurate order) of such magnetic textures can often be quite large (and unlimited for incommensurate order) which currently makes {\it accurate} first-principle descriptions challenging if not prohibitive.

An often adopted strategy is to turn to spin model Hamiltonians, with 
parameters extracted for first-principle calculations. This considerably simplifies the problem,
but without necessarily making it easily solvable.  A case in point is provided  by 
the quantum Heisenberg model (QHM), the model in focus in this work: Exact numerical methods
like exact diagonalization(ED), DMRG, or Monte Carlo can be applied for not too large samples, and highly valuable 
information extracted for ferro-, antiferro, ferri-magnetic orders. 

However, for large number of atoms/spins in the magnetic pattern, a significant increase in size is needed, 
(as, e.g., for dipolar interactions, or when antisymmetric
exchange couplings are present); here, using spatial/spin symmetries, the computational difficulty can be mildly reduced,
but not eliminated.  As a concrete example, for magnetic skyrmions it is the anti-symmetric exchange that leads to a novel spin texture. For this type of exchange coupling,, the total spin-$z$ operator does not commute with the Hamiltonian, and 
configurations with different spin-$z$ are mixed. Furthermore, skyrmion spin-textures extend over several lattice distances, and direct exact numerical methods become inadequate at these sizes. 

The alternative considered in this work is the Green's function (GF) formalism that, even for considerable 
lattice/cluster sizes, remains quite affordable from the computational point of view. In this paper, we have derived the magnon Hedin's equations via the Schwinger functional derivative technique, and applied the GF method to 
solve Heisenberg model on 2D lattices, comparing the results with ED benchmarks.

On the one hand, for a $J_1-J_2$ model on square and hexagonal lattices the GF scheme beyond the RPA gives reasonable accuracy with relatively low computational costs (in our comparisons, the limiting factor for the size of the clusters was in fact the ED treatment). The considerations apply to both ground state spin patterns and to magnetic structure factor results. Furthermore, th same level of agreement was found for inhomogeneous clusters in the few-impurity limit.

On the other hand, a preliminary attempt to apply the GF scheme to systems with an antisymmetric exchange interaction was not equally satisfactory. Our results suggest that a cross-product spin coupling may require a better approximate 
prescription than the neglect of the high-order response of the GF to the probing field. Possible procedure for improvements in this direction are currently being explored. 

Finally, based on the outcomes of this work, we expect that the current approach (and the inherent approximation scheme used here) will work for long-range scalar product (e.g. magnetic-dipole) interactions. As another follow-up of this work, we plan to 
investigate systems with magnetic dipole interactions and compare the GF results with experimental values. Quite naturally, the long term goal is to improve the GF scheme in a way that it would become possible to accurately calculate 
the magnetic structure factor of real materials.

\section{Acknowledgments}
F.A. gratefully acknowledges
financial support from the Knut and Alice Wallenberg  
Foundation (KAW 2017.0061) and the Swedish Research Council 
(Vetenskapsrådet, VR 2021-04498\_3).

\bibliography{bib}
\bibliographystyle{unsrt} 

\end{document}